\documentclass[12pt,notoc,article]{JHEP3}
\usepackage{amsmath,amssymb,euscript,array,mathrsfs,epsfig}
\def\ben
{\begin{equation}}
\def\een{\end{equation}}

   \let\d=\delta

\def\be{\begin{equation}}
\def\ee{\end{equation}}
\def\ba{\begin{array}}
\def\ea{\end{array}}

\def\dalemb#1#2{{\vbox{\hrule height .#2pt
        \hbox{\vrule width.#2pt height#1pt \kern#1pt
                \vrule width.#2pt}
        \hrule height.#2pt}}}

\newcommand{\bea}{\begin{eqnarray}}
\newcommand{\eea}{\end{eqnarray}}

\newcommand{\Tr}{{\rm Tr} }

\title{Heavy quark density in ${\cal N}=4$ SYM: from hedgehog to Lifshitz spacetimes}
\author{$\quad$ S. Prem Kumar\\\\
{\it 
Department of Physics, \\
Swansea University, \\ 
Singleton Park, Swansea\\
SA2 8PP, U.K.  
}\\
E-mail: \email{s.p.kumar@swansea.ac.uk}
} 
\abstract{
We study the effect of an ${\cal O}(N^2)$ density of heavy quarks in strongly coupled ${\cal N}=4$ SUSY Yang-Mills theory in the large $N$ limit. This is achieved in the type IIB supergravity dual by introducing a uniformly smeared density of macroscopic string sources stretching to the boundary of $AdS_5\times S^5$. The backreacted system exhibits a flow from an $AdS_5$ ``hedgehog'' geometry to a scaling Lifshitz-like solution ${\rm Lif}_{\rm 5}\times S^5$ with dynamical critical exponent $z=7$, wherein the scaling symmetry is broken by a logarithmic running dilaton. We find an exact black brane solution within the scaling regime which describes the low temperature thermodynamics of the system.
}
\begin{document}
\section{Introduction}
 The AdS/CFT correspondence \cite{maldacena, witten} has been instrumental in unravelling many facets of certain types of strongly interacting field theories in the large-$N$ limit. Within this framework, the behaviour of matter at finite density has been intensely studied with a view towards applications in condensed matter physics (see reviews \cite{sean, herzog, mcgreevy} and references therein). The physics of cold, dense QCD matter 
presents equally exciting challenges \cite{rajagopal, schafer, stephanov} for holographic approaches involving flavour branes in both the probe limit \cite{Karch:2002sh, mateos, andy, ghoroku} and incorporating their backreaction \cite{aldo}.

The focus of this paper will be to study the effect of a sizeable, spatially homogeneous number density of heavy quark ``impurities" introduced as external sources into ${\cal N}=4$ supersymmetric Yang-Mills theory in the large-$N$ limit and at strong 't Hooft coupling.  The number density of such heavy quarks will be chosen to scale as $N^2$, so that we may then study the backreaction of these on the gauge theory. One of the motivations for considering this setup is to obtain intuition for the finite density thermodynamics of holographic models with a large number of flavours (with $N_f/N$ fixed in the large-$N$ limit).

According to the dictionary provided by gauge/string duality, a heavy quark transforming in the fundamental representation of the $SU(N)$ gauge group corresponds to a macroscopic fundamental string stretching to the boundary of the dual geometry \cite{malwil,reyyee}. The end-point of the string is anchored to the worldline of the quark in the boundary gauge theory. For a single quark, it suffices to treat the string as a probe in the dual geometry. It is natural to ask what happens when a large number of such quarks is introduced and, in particular, when this number scales as $N^2$ in the large-$N$ limit, so that the quark impurities necessarily backreact on the gauge degrees of freedom. This is a difficult question to answer in general, but it is conceivable that a configuration preserving sufficient amount of global symmetries may render the problem tractable within the dual gravity description at strong coupling. 

For a single heavy quark, or a straight timelike Wilson line, the corresponding string world-sheet is localized at a point on the transverse $S^5$ in $AdS_5\times S^5$, breaking the $SO(6)$ global symmetry to $SO(5)$. If we consider a large number of such string sources, we may distribute them uniformly along all transverse directions, thereby restoring both spatial translational invariance and the $SO(6)$ global symmetry in the boundary theory. The energy density  associated to such smeared distributions \footnote{Recently, smeared string configurations have appeared in the study of finite density physics in a different system \cite{Polchinski:2012nh}.} scales as $\sim n_{\rm F1}/(2\pi\alpha') = n_{\rm F1}\,\sqrt{\lambda}/2\pi$, where $n_{\rm F1}$ is the number density of macroscopic strings or heavy quarks and $\lambda$, the (large) 't Hooft coupling. The heavy quark distribution then backreacts on the degrees of freedom of the ${\cal N}=4$ theory when $n_{\rm F1} \sim N^2/\sqrt\lambda$. 

One might worry that a smearing of string sources along all transverse directions would lead to violation of Gauss' law applied to the sources. Indeed, the Chern-Simons couplings of IIB supergavity require that such smearing is consistent provided that a certain three-form flux is also switched on. Intuitively, the resulting three-form flux may be interpreted as a density of baryon vertices 
\cite{baryons} which soak up the charge corresponding to the string sources.

Our primary goal in this work is to understand the zero/low temperature behaviour of this highly symmetric smeared system as a simplified model for a finite density state in a theory with a known string dual. 
We further restrict attention to the ${\cal N}=4$ theory on ${\mathbb R}^{3,1}$. We postpone the study of the thermodynamics of the model, and specifically its finite volume phase structure,  to future work
\cite{inprog}. Since the quarks we consider are infinitely heavy, quark number cannot fluctuate and therefore all thermodynamic questions are restricted to the canonical ensemble. Related to this point is the absence of a global $U(1)_B$ baryon number symmetry in this system. However, there is a natural embedding of the above configuration into a holographic model with dynamical massive quarks and a $U(1)_B$ symmetry - namely the D3/D7 system with a large number of smeared flavour branes \cite{aldo, Nunez:2010sf}.

The main result of this paper is that the backreaction of the heavy quark density on ${\cal N}=4$ theory triggers a flow to an infrared regime which exhibits approximate dynamical or Lifshitz scaling \cite{Kachru:2008yh} i.e. invariance under the scale transformations $t\to a^z\, t$ and $\vec x\,\to a\,\vec x$. The dynamical critical exponent for the system under investigation has a fixed value, $z=7$, independent of any free parameters
\footnote{After this paper was released, we were made aware by the authors of \cite{Azeyanagi:2009pr} that this particular scaling solution has appeared in that paper (see Appendix C of \cite{Azeyanagi:2009pr}).}
. Importantly, the Lifshitz scaling symmetry is only approximate, due to logarithmic running of the dilaton in the gravity dual. This behaviour closely resembles the scaling solutions found in \cite{Taylor:2008tg, Goldstein:2009cv, Goldstein:2010aw, Charmousis:2010zz} for charged dilaton black branes. As in the latter systems, the running of the dilaton renders $\alpha'$ corrections important deep in the infrared (IR). By electric-magnetic duality of the ${\cal N}=4$ theory, and S-duality of type IIB string theory, the same statements apply to magnetic sources (replacing the strings by D1-branes), except that now the dilaton becomes large in the deep infrared, necessitating the inclusion of string loop corrections.
 
We also find an exact black brane solution in the Lifshitz-like scaling regime, which, by virtue of a small non-vanishing temperature, shields the system from the regime where the classical supergravity approximation breaks down. Taken at face value this solution has vanishing entropy density in the zero temperature limit, a physically appealing feature. However, clearly, in the strict zero temperature limit the corrections mentioned above will alter the conclusions somewhat (e.g. \cite{Harrison:2012vy}). It is nevertheless extremely interesting to learn that the effect of a finite heavy quark density is qualitatively different to the effect of a finite R-charge density in the ${\cal N}=4$ theory. As is well known, the latter, for a specific choice of R-symmetry chemical potentials yields the Reissner-Nordstrom solution \cite{chamblin} that has a large entropy density at zero temperature. It is worth noting that the emergence of Lifshitz-like scaling as a result of backreacting distributions of D-branes/baryons has also been seen generally in
\cite{Hartnoll:2009ns}.

It should be pointed out that the backreaction of heavy quarks or Wilson lines on ${\cal N}=4$ SYM has been studied in two different contexts. The first of these are the bubbling geometries dual to supersymmetric half-BPS Wilson lines in large representations (with ranks of order $N^2$) \cite{Lunin, DHoker}. The second context which is closely related to the discussion in this paper is the work of Headrick in \cite{headrick} where the effect of smeared Polyakov loops on ${\cal N}=4$ SYM at finite temperature was investigated.

This paper is organized as follows. In Section 2, we derive the reduced 5D action describing the smeared system of strings. The features of the ultraviolet behaviour of the resulting solutions are discussed and clarified in Section 3. Sections 4 and 5 are respectively devoted to the derivation of the IR scaling solution and the (numerical) construction of the flow from $AdS_5\times S^5$ to the Lifshitz-like scaling regime. We discuss possible future directions in the final section.

\section{IIB plus strings}

As explained above, we take a finite heavy quark density in the field theory to correspond to a geometry sourced by a uniform distribution of fundamental strings, each stretching to the conformal boundary of $AdS_5\times S^5$. For simplicity the strings are chosen to be smeared uniformly along the internal $S^5$ directions. This preserves the $SO(6)$ global R-symmetry  of ${\cal N}=4$ SYM, but breaks all supersymmetry.
 The coupled system of type IIB supergravity and a uniform density of macroscopic fundamental strings is described by the action 
\be
S \, =\, S_{\rm IIB}\, + \,S_{\rm F1}\,+\,
\frac{1}{2\pi\alpha'}\, n_{\rm F1}\,
\int\,B_{2}\wedge dx^1\wedge dx^2\wedge dx^3\wedge \omega_5\,\frac{1}{\pi^3}\,.
\ee
Here $S_{\rm F1}$ is the Nambu-Goto action for the string sources, $n_{\rm F1}$ is the number density of heavy quarks (or strings) in the boundary theory, $\omega_5$ the volume form on the unit five-sphere while $B_2$ is the Neveu-Schwarz two-form potential which couples to the string world-sheet . 

We then look for a consistent solution to the supergravity equations which preserves $SO(6)$ global symmetry, the spatial translational and rotational invariance, and  with the correct fluxes turned on to account for the presence of a non-vanishing fundamental string charge. 
The relevant field equations for the three-form and five-form fluxes
$H_3, F_3$ and $\tilde F_5\equiv F_5 + \tfrac{1}{2}B_2\wedge F_3 -\tfrac{1}{2}H_3\wedge C_2
$ in Einstein frame, are:
\bea
&& d *\tilde F_5\, =\, H_3\wedge F_3\,,\qquad\qquad d\,(e^{\hat\phi} *F_3)\, =\, - 
g_s\,H_3\wedge F_5\,,\\\nonumber\\\nonumber
&& g_s \,d\, (e^{-\hat\phi} *H_3)\, =\, - 
g_s^2\, F_5\wedge F_3\, + \,
\frac{n_{\rm F1}}{2\pi\alpha'}\,\frac{16\pi G_N}{\pi^3}\,
\omega_{5}\wedge dx^1\wedge dx^2\wedge dx^3\,.
\eea
The asymptotic value of the dilaton $e^{\hat\phi}$ is set to $g_s$ the string coupling, while $G_N$ is Newton's constant in ten dimensions, with $16\pi G_N\,=\,(2\pi)^7\,g_s^2\,\alpha^{\prime\,4}$\,.
The field equations are then solved by 
\bea 
&&g_s\,F_5 
= \,4\,(1+*)\, \omega_{5}\,,\qquad\qquad H_3=0\,,
\label{eff5}\\\nonumber\\\nonumber
&& F_3\,=\,\frac{n_{F1}}{N} \,\frac{4\pi^2}{\sqrt\lambda}\,dx^1\wedge dx^2\wedge dx^3\,.
\eea
We have used the standard AdS/CFT dictionary, namely that the 't Hooft coupling 
$\lambda= R^4_{\rm AdS}/\alpha'^2 = (4\pi g_s N)$, and we have conveniently  rescaled the AdS radius $R_{\rm AdS}$ to unity. The introduction of macroscopic fundamental strings stretching from the boundary of $AdS_5$  to the origin, necessarily requires a non-zero three-form field strength. 

A non-vanishing flux for $F_3$ would suggest the presence of D5-brane sources. Although there were no explicit D5-branes in the setup to  begin with, we could interpret the source of the three-form flux as a distribution of D5-branes wrapped on $S^5$ with a number density, $n_5 \sim \frac{n_{\rm F1}}{N}$. This is consistent with the well known fact that $N$ quarks must be bound into a gauge-invariant baryon operator which corresponds to a wrapped D5-brane with $N$ strings attached \cite{baryons}.
Below we want to obtain the backreacted geometry  wherein the fundamental string distribution and associated baryons can be replaced with smooth backgrounds accompanied by corresponding fluxes, and no additional D-branes. Since the fluxes are completely determined as above in terms of $n_{\rm F1}$ and $N$, the only `active' fields in the background are the dilaton and the metric components.

\subsection{Metric Ansatz}
Given the symmetries of the problem, the metric and the dilaton  have the form
\bea
&&ds^2 = g_{rr}(r)\,dr^2 + g_{tt}(r)\, dt^2 + e^{2\sigma(r)}\,d\vec x^{\,2}+
e^{2 \eta(r)}\,d\Omega_5^2\,,\\\nonumber
&&e^\phi\,\equiv\, e^{\hat \phi}\, g_s^{-1}\,,
\eea
where we have defined the shifted dilaton $\phi$ which vanishes at infinity. In this background, the Nambu-Goto action for the smeared distribution of strings is,
\be
S_{\rm F1}\,=\,\frac{n_{\rm F1}\,\sqrt\lambda}{2\pi} \,\int d^4x\,\int dr\,\sqrt{- g_{rr} \,g_{tt}}\,e^{\phi/2}\,.
\ee
This is accompanied by a term quadratic in $n_{\rm F1}$, arising from the energy density in $F_3$.
We will find it useful to work with the 5D effective action after appropriate rescaling of the 10D metric:
\bea
&&\hat g_{\alpha\beta}\,=\,e^{10\,\eta/3}\,g_{\alpha\beta}\,,\qquad \alpha,\beta\,=\,0,1,\ldots 4\,.\\\nonumber\\\nonumber
&& ds^2_{(5)}\,\equiv\,\hat g_{\alpha\beta}\,dx^\alpha dx^\beta\,=\,\hat g_{tt}\,\,dt^2\,+\,\hat g_{rr}\,\,dr^2\,+\,e^{2\hat\sigma}
\,\,d\vec x^2\,,\qquad\hat\sigma\,\equiv\,\sigma+\tfrac{5}{3}\eta\,.
\eea
This ansatz along with eq.\eqref{eff5}, when substituted into the type IIB supergravity action,  yields the reduced 5D system:
\bea
&&S\,=\,-\frac{N^2}{8\pi^2}\,\int\,d^5x\,\sqrt{\hat g}\,
 \left(\hat R\,-\tfrac{40}{3}\,\eta^{\prime\,2}\,\hat g^{\,rr}\,-\,\tfrac{1}{2}\,\phi^{\prime\,2}\,\hat g^{\,rr}+\right.
 \label{5D}\\\nonumber
&&\left.\hspace{1.8in}+\,20\,e^{-16\eta/3}-\tfrac{1}{2}\,\left(Q\,e^{\phi/2}\,e^{10\eta/3}\,e^{-3\hat\sigma}\,+\,4\,e^{-20\eta/3}\right)^2\right)\,.
 \eea
The first of the potential energy terms is the scalar curvature of the $S^5$. The contributions from the $F_3$ and $F_5$ fluxes, and the smeared Nambu-Goto action for the strings can be packaged neatly as a perfect square
\footnote{Since this action actually describes  a one-dimensional (radial) system, it may be possible to obtain a superpotential for it to explain the perfect square structure, and to derive first order equations. I thank A. Faedo for drawing attention to this point.}. The bulk action depends on the quark number density via a single parameter $Q$ defined as
\be
Q\,\equiv\, \sqrt{\lambda}\,\pi \,\frac{n_{\rm F1}}{N^2} \,=\,{\sqrt\lambda}\,\pi\,\frac{n_5}{N}\,.
\ee
$Q$ is a density and therefore a dimensionful parameter. With regard to its dependence on the number of colours $N$, we will treat it as a number of order one in the large $N$ limit, so that the quark density is ${\cal O}(N^2)$, or equivalently, the baryon number density $n_5$ is ${\cal O}(N)$. 
The system we are studying and the form of the action above is closely related to that of \cite{headrick} which did not need to include a non-zero $F_3$ 
\footnote{In \cite{headrick}, the backreaction of a uniformly smeared configuration of strings connecting {\em antipodal} points of the boundary $S^3$ was analyzed (in global $AdS_5$).}.

\section{UV AdS-hedgehog solution}

The backreaction of a uniform distribution of string sources on 
pure gravity (in four dimensions) was studied in \cite{gundelman} and termed a ``hedgehog'' black hole. This was then generalized to asymptotically AdS spacetimes in \cite{headrick}. The characteristic feature of the so-called hedgehog configurations in $AdS_5$ is that the gravitational potential includes a term $\sim  -\,Q/r$ for large radial coordinate $r$, directly following from the linear dependence of the mass of a stretched string on its length.

In the present case, in the gauge $\hat g_{rr} =\frac{1}{r^2}$,  we obtain the following large $r$ asymptotic expansions for the metric functions and the dilaton,
\bea
\hat g_{rr}\,=\,\frac{1}{r^2}\,,\qquad
&&\hat g_{tt}\,=\,r^2\,-\,\frac{8\, Q}{9\, r}-\frac{\varepsilon}{r^2} +\frac{39233}{158760}\,\frac{Q^2}{r^4}\,+\ldots\,,\label{uv}\\\nonumber\\\nonumber
&& e^{2\,\hat\sigma}\,=\, r^2+ \frac{4\, Q}{9\, r}+\frac{\varepsilon}{3\, r^2}
-\frac{877}{158760}\,\frac{Q^2}{r^4}\,+\ldots\,,\\\nonumber\\\nonumber
&& e^\phi\,=\,1-\frac{2\, Q}{3\, r^3} + \,\frac{v_4}{r^4}\,+\frac{101}{504}\,\frac{Q^2}{r^6}\,+\ldots\,,\\\nonumber\\\nonumber
&&e^{2\,\eta}\,=\,1+\frac{Q}{35\, r^3}- \frac{5441}{98000}\,\frac{Q^2}{r^6}+\frac{v_4-\varepsilon}{22}\,\frac{Q}{r^7}\,+\,\frac{v_8}{r^8}\,+\ldots
\eea
We note that the leading correction to the $AdS_5$ metric scales as $\sim - Q/r$, which is due to the macroscopic string source in the bulk. This breaks conformal invariance and Lorentz invariance of the boundary theory and the metric is not strictly asymptotically AdS. However it is of the asymptotically locally AdS (AlAdS) form \cite{yiannis} {i.e.} for large $r$,
\be
\hat R_{\mu\nu\lambda\sigma}\,=\,(\hat g_{\mu\sigma}\,\hat g_{\nu\lambda}-\hat g_{\mu\lambda}\,\hat g_{\nu\sigma})\,\left(1+{\cal O}\left(\tfrac{Q}{r^3}\right)\right).
\ee
After fixing the asymptotic form of the metric, the solution space is characterized by three integration constants $\varepsilon, \,v_4$ and $v_8$ corresponding to normalizable modes. The active fields in the solution are the dilaton $\phi$, the metric component $g_{tt}$, and the $s$-wave components of the volume scalars $\hat\sigma,\,\eta$ of the ${\mathbb R}^3$ and $S^5$ respectively. The volume scalar $\eta$ is dual to the irrelevant, dimension eight operator ${{\cal O}_8}\,=\,\tfrac{1}{N}{\rm STr}\,\left[F^4-\tfrac{1}{4}\,(F^2)^2\right]$,  where `STr' denotes the symmetrized trace \cite{ferrara, tseytlin}. On the other hand the dilaton is dual to the marginal operator ${\cal O}_4\,=\,\tfrac{1}{N}{\rm Tr} F^2$.  The integration constants $v_4$ and $v_8$ are the expectation values of ${\cal O}_4$ and ${\cal O}_8$ respectively, whilst $\varepsilon$ is related to the energy density of the boundary field theory. Note that we have taken the value  of the dilaton $\phi$ to vanish asymptotically. In principle this is a tunable (dimensionless) parameter in the solution corresponding to the marginal gauge coupling of ${\cal N}=4$ theory.

In summary, the heavy quark density in the boundary gauge theory does not introduce irrelevant operators/non-normalizable modes in the bulk. It does mildly alter the asymptotics. In the absence of dimensionful VEVs $v_4$, $v_8$ and $\varepsilon$, the background can be obtained as an expansion in powers of $\frac{Q}{r^3}$ since $Q$ is a number density and has canonical scaling dimension three (in the UV). In general, we have no reason to expect the VEVs  to vanish. The existence of a smooth infrared (IR) solution may well require the dimensionful VEVs to be turned on (both at zero and non-zero temperatures). The possible values of these can only be ascertained after the IR dynamics of the finite density state and the complete flow towards it is understood. 

Before turning to the putative IR behaviour of the model under consideration, we briefly touch upon features of planar AdS-hedgehog geometries in Einstein gravity and how holographic regularization of the action of so-called hedgehog configurations proceeds.

\subsection{Holographic regularization} 

A potential subtlety associated with the background above concerns the definition of a properly renormalized bulk action. This is an issue since the background contains sources which generate corrections ($\sim -Q/r$) to the bulk $AdS_5$ metric.  To clarify this, let us first look at the divergent terms in the full 5D bulk (Euclidean) action, obtained by plugging in the IIB hedgehog asymptotics \eqref{uv},
\be
S_{\rm bulk}\,=\,-\frac{N^2}{8\pi^2}\int d^4x\, \int_{r_h}^{\Lambda} dr \, ( - 8\,r^3 - \tfrac{28}{9}\, Q+\ldots)\,,
\ee
where ${\Lambda}$ is the UV cut off, which we treat as the boundary. 
Following the standard approach towards holographic renormalization 
\cite{hennskend,ejm,vijaykraus},  we must also include the Gibbons-Hawking (GH) boundary term and possible geometrical counterterms. In the situation with a planar boundary the only non-vanishing counterterm arises from the boundary cosmological constant. So we obtain, 
\bea
&& S_{\rm GH}\,=\, - \frac{N^2}{8\pi^2}\int d^4x\,\sqrt{h}\,\Tr\, K\,\big |_{r=\Lambda}\,=\,- \frac{N^2}{8\pi^2}\int d^4x\,\left(\, 8\, \Lambda^4+\tfrac{4}{9}\,Q\,\Lambda+\ldots\right)\\\nonumber\\\nonumber
&& S_{\rm ct}\,=\, \frac{N^2}{8\pi^2}\int d^4x\,\sqrt{h}\,6\big |_{r=\Lambda}\, =\,
\frac{N^2}{8\pi^2}\int d^4x\,\left(\, 6\, \Lambda^4+\tfrac{4}{3}\,Q\,\Lambda+\ldots\right)\,.
\eea
Here $h_{\alpha\beta}$ and $K$ are  the induced metric  and the extrinsic curvature of the boundary at $r=\Lambda$, respectively. Summing the bulk and boundary actions yields
\be
S_{\rm bulk}+S_{\rm GH} + S_{\rm ct}\,=\, \frac{N^2}{8\pi^2}\,\int d^4 x\,\, 4 \,Q\,\Lambda+\ldots
\ee
The usual boundary terms are sufficient to cancel the leading divergences, leaving behind a linear divergence proportional to the number density of strings/quarks. We will now see that this term can also be removed by adding an appropriate counterterm for the source action, namely a boundary term for the Nambu-Goto action of the macroscopic strings that source the geometry. In particular, a string stretching to the boundary of AdS space has to satisfy a Neumann boundary condition along the radial direction, and four Dirichlet conditions along the directions parallel to the boundary. To this end, we need to include a term in the F-string action which replaces the radial coordinate with its conjugate momentum
\cite{dgo, drukkerfiol},
\be
S_{\rm F1} \to S_{\rm F1} +\delta S_{\rm F1}\,=\, S_{\rm F1} - \int dt \,r\, \frac{\delta S_{\rm F1}}{\delta r'}\big|_{r=\Lambda}\,.
\ee
Here $r'$ denotes a derivative with respect to the world-sheet spatial coordinate. For the smeared configuration of strings this is precisely given by 
\be
\delta S_{\rm F1} \,=\, - \frac{N^2}{8\pi^2}\,\int d^4x\,\, 4\,Q\,\Lambda\,e^{\phi/2-10\eta/3}\,\sqrt{\hat g_{rr}\,\hat g_{tt}}\,\big|_{r=\Lambda}\,,\label{ct2}
\ee
which cancels the linear divergence of the bulk action. This 
divergence\footnote{ The prefactor $\tfrac{N^2}{8\pi^2}\times 4 Q$, in eq.\eqref{ct2} is equal to $\frac{n_{\rm F1}}{2\pi\alpha'}$.} has a simple physical interpretation -- it corresponds to the bare mass of the heavy quark, which is formally infinite and must naturally be subtracted to define a sensible energy density.

\subsection{AdS-hedgehog black hole in pure gravity}

In pure gravity in five dimensions with a negative cosmological constant, the backreaction of a uniform distribution of strings gives rise to a background with metric (in Schwarzschild-like  coordinates),
\bea
ds^2\,=\,-f(\tilde r)\,dt^2 +\frac{d\tilde r^2}{f(\tilde r)} \,+\,\tilde r^2\,d\vec x^{\,2}\,,\qquad f(\tilde r)\,=\, \tilde r^2 - \frac{4\, Q}{3\, \tilde r}- \frac{r_h^4-\tfrac{4}{3}Q\, r_h}{\tilde r^2}\,.
\eea
This has a horizon at $\tilde r=r_h$.
In a Fefferman-Graham type expansion (where the boundary is at $r\to\infty$), the metric can be recast as
\bea
&&ds^2\,=\, \frac{dr^2}{r^2}\,+\,\left(r^2 -\frac{8\,Q}{9\, r}-\frac{\varepsilon}{r}+\ldots\right)\, dt^2 + \left(r^2+\frac{4\, Q}{9\, r}+\frac{\varepsilon}{3\, r}+\ldots\right)\,d\vec x^{\, 2}\,,\\\nonumber
&&\varepsilon\,=\, \tfrac{3}{4}r_h^4 - Q\, r_h\,,
\eea
which agrees with the asymptotic behaviour eq.\eqref{uv}. The thermodynamics of such solutions is non-trivial, particularly when the horizon is spherical \cite{headrick}. In the planar case, it is easily seen that the hedgehog black hole in GR has a smooth zero temperature  limit wherein $f(\tilde r)$ has a double zero at $\tilde r = r_h$, resulting in a non-vanishing entropy density $s$,
\be
s\,=\, \frac{N^2}{2\pi}\, \frac{Q}{3}\,,\qquad\qquad
r_h\,\equiv\,\left(\frac{Q}{3}\right)^{\tfrac{1}{3}}\,.
\ee
Thus the zero temperature system (in Einstein gravity) has a non-zero, large entropy density and the IR physics is described by the near horizon $AdS_2\times {\mathbb R}^3$ geometry. This feature, similar to the Reissner-Nordstrom black hole, seems unwanted and we may expect the type IIB setup dual to ${\cal N}=4$ SYM with heavy quark density to have a different and physically more appealing IR behaviour.

\section{Exact (IR) scaling solution} 
The most interesting question about the system we are studying is with respect to its low energy, long wavelength behaviour at low/zero temperature. We have seen that, in the pure gravity case, the low energy physics is dictated by a near horizon $AdS_2$ region. We would now like to understand whether the type IIB hedgehog has a substantially different IR behaviour leading to a zero entropy ground state at zero temperature.

We find that the system of ``IIB plus strings'' also admits an exact scaling solution. It was not {\it a priori} clear that the system described by eq.\eqref{5D} should possess such a solution. In fact, one may have naively expected the UV AdS-hedgehog geometry to flow in the IR to a background sourced by multiple D5-branes wrapped on $S^5$ (corresponding to baryon vertices located at the origin of the space). However, closer inspection reveals no clear separation of scales (in terms of the radial coordinate $r$) wherein the energy density in the three-form flux $\sim F_3^2$ dominates over the cosmological constant and the string density. This leads us to look for scaling solutions where all terms in the potential are comparable. We find, using the same coordinate system as in the UV hedgehog asymptotics, namely with $\hat g_{rr}\,=\,1/r^2 $, an {\em exact} scaling solution to the equations of motion,
\bea
&&ds^2_{(5)}\,=\, \frac{dr^2}{r^2}\,-\,r^{14/\ell}\,dt^2\,+\,r^{2/\ell}\,d\vec x^{\,2}\,,\qquad\qquad \ell\,=\,\sqrt{10}\,\left(\frac{136}{121}\right)^{1/3}\,\,,\label{sol1}\\\nonumber\\\nonumber
&& e^{\phi}\,=\, \frac{11979}{578\,\sqrt{34}\,\, Q^2}\,r^{6/\ell}\,,\qquad\qquad e^{2\eta}\,=\,\left(\frac{136}{121}\right)^{1/4}\,\simeq\,1.0297\,.
\eea
This is a constant curvature spacetime, which we can cast in a more conventional form by defining a new radial coordinate $\rho\,\equiv r^{1/\ell}$, and after a rescaling
\bea
&& ds^2_{(5)}\,=\,\ell^2\left(\frac{d\rho^2}{\rho^2}\,+\, \rho^2\,d\vec x^{\,2}\,-\,\rho^{14}\,dt^2\right)\,,\label{sol2}\\\nonumber\\\nonumber
&&e^\phi\,=\,\sqrt{\frac{2}{17}}\,\frac{144000}{11 \, Q^2}\,\rho^{6}\,.
\eea
Therefore the scaling solution describes a regime with Lifshitz-like anisotropic scaling \cite{Kachru:2008yh} and dynamical critical exponent $z=7$. The scaling symmetry is not exact due to the logarithmic running of the dilaton with the radial coordinate $\rho$.

The first observation is that the  dynamical exponent is independent of the heavy quark density $Q$.  
It is also clear that the full ten dimensional metric is smooth in Einstein frame, since the $S^5$ has a constant radius $e^\eta = \sqrt[8]{\frac{136}{121}}\simeq 1.015$\,, while the Ricci scalar for the Lifshitz metric with $z=7$ is, 
\be
\hat R\,_{(5)}=\, - 2\,(z^2\,+\,3\,z\,+\,6)\,\ell^{-2}\simeq -14.06\,.
\ee
However, as the coupling $e^\phi$ runs to zero at $\rho=0$, the string frame metric has a curvature singularity in the deep infrared and the solution cannot be trusted since it will receive large $\alpha'$ corrections
\footnote{Note that this string frame curvature singularity is completely distinct from the effect of divergent tidal forces of backgrounds with exact Lifshitz scaling \cite{Kachru:2008yh,Horowitz:2011gh}}. Nevertheless, as we will see, we can always introduce a temperature into the system and cloak the singularity behind a horizon, without spoiling the scaling regime and therefore steer clear of the potential problems near $\rho=0$. In this respect the solution \eqref{sol2} is quite similar to the charged dilaton black brane metrics of 
\cite{Taylor:2008tg, Goldstein:2009cv, Goldstein:2010aw}. 

Since the dilaton grows without bound for large $r$, in order to make sense of this solution we must be able to embed the scaling regime within the flow originating from the asymptotically (locally) AdS hedgehog background.

It is worth noting that any putative baryon vertices  (wrapped D5-branes) which could be regarded as sources of the $F_3$ flux, are absent, possibly hidden by the ``Lifshitz horizon''. Hence, for any small non-zero temperature, the system of heavy quarks is in a deconfined phase. Within the type IIB framework we can act on our system of smeared fundamental strings with the SL(2,${\mathbb Z}$) duality group. The action of S-duality, for example, replaces the F-strings with D-strings (corresponding to magnetic sources on the boundary) and simply changes the sign of the dilaton, so that $e^\phi$ diverges at $\rho=0$, requiring the inclusion of quantum (loop) corrections in the deep IR.

\subsection{Effective Potential}
Let us now briefly explain the appearance of the Lifshitz solution by examining the equations of motion. The UV AdS-hedgehog solution nominally depends on two parameters: $Q$ and the asymptotic value of the dilaton. However, in the absence of a scale in ${\cal N}=4$ SYM, bulk gravity solutions with different values of $Q$ are related by a straightforward rescaling of the radial coordinate, and therefore do not really represent new solutions. On the other hand, the potential in the 5D action \eqref{5D} depends on the combination $Q^2\, e^\phi$, so that UV solutions with different asymptotic values of the dilaton are related by a rescaling of $Q$. Given this, we may  distinguish different UV solutions by the asymptotic value for the parameter $Q^2\, e^\phi$. Independently of this asymptotic value, in the scaling Lifshitz-like solution eq.\eqref{sol2}, the combination $Q^2\, e^\phi$ is completely fixed and there are no free parameters (apart from coordinate rescalings). This indicates an attractor behaviour  along the lines of what was found in \cite{Goldstein:2009cv, Goldstein:2010aw}. 

It is perhaps easiest to analyze the equations of motion in the gauge 
$\hat g_{tt} = 1/\hat g_{\varrho\varrho}$ where $\varrho$ is an appropriate radial coordinate. From eq.\eqref{5D}, one identifies the natural effective potential for the one-dimensional (radial) problem,
\be
V\,=\,\frac{e^{3\hat\sigma}}{2}\,\left(Q\,e^{\phi/2}\,e^{10\eta/3}\,e^{-3\hat\sigma}\,+\,4\,e^{-20\eta/3}\right)^2- 20\,e^{3\hat\sigma-16\eta/3}\,.
\ee
The equations of motion (one of which is a constraint) in this gauge
are,
\bea
&&\left(\hat g_{tt}\left(e^{3\hat\sigma}\right)^\prime\right)^\prime\,+\, V\,=\,0\,,\qquad\qquad\left(\hat g_{tt}\,e^{3\hat\sigma}\,\eta'\right)^\prime\,-\, 
\tfrac{3}{80}\,\partial_\eta V\,=\,0\,,\\\nonumber\\\nonumber
&&\left(\hat g_{tt}\,e^{3\hat\sigma}\,\phi'\right)^\prime\,-\, 
\partial_\phi V\,=\,0\,,\qquad\qquad
3\,e^{-\hat\sigma}\,\left(e^{\hat\sigma}\right)^{\prime\prime}\,
+\,\tfrac{1}{2}\,\phi^{\prime\,2}\,+\,\tfrac{40}{3}\,\eta^{\prime\,2}
\,=\,0\,.
\eea
The effective potential has a runaway behaviour, with a critical point at infinity where the dilaton $e^\phi$ and the volume scalar 
$e^{\hat\sigma}$ vanish (and so does the potential itself). The equations of motion then admit a scaling solution where this critical point is reached at the horizon ($\varrho=0$) by logarithmic running of $\phi$ and $\hat\sigma$. In this scaling regime the scalars $\eta$ and the combination 
$ e^{\phi-6\hat\sigma}$, both remain fixed at constant values. It is a straightforward excercise to check in this gauge that the field equations are solved by
\bea
&&\hat g_{tt}\,=\,\frac{7^2}{\ell^2}\,\varrho^2\,,
\qquad\qquad e^\eta\,=\,\sqrt[8]{\tfrac{136}{121}}\,,
\\\nonumber\\\nonumber
&&e^{\hat\sigma}\,=\,c_0\,{\varrho}^{1/7}\,,\qquad\qquad
e^\phi\,=\,c_0^{6}\,\tfrac{11979}{578\,\sqrt{34}\,Q^2}\,{\varrho}^{6/7}\,,
\eea
where $\ell$ is defined in eq.\eqref{sol1}
The individual normalizations of $e^{\hat\sigma}$ and $e^\phi$ can be changed by a coordinate rescaling subject to the requirement that the combination $e^{\phi-6\hat\sigma}$ is fixed. After a coordinate transformation the background can be put in the Lifshitz-like form \eqref{sol2}

\subsection{Black hole in scaling regime}

We have already pointed out that the dilaton running to zero in the infrared implies large string frame curvatures in the deep IR, requiring the inclusion of $\alpha'$ corrections in the gravity dual. Introducing a finite temperature would shield the system from this regime where the gravity description becomes inaccurate, and provide a consistent gravity dual of the boundary field theory at strong coupling.  We find that the field equations admit an exact solution which is a simple deformation of the Lifshitz background, yielding the following metric with a horizon in Schwarzschild-like coordinates:
\bea
&& ds^2_{(5)}\,=\,-\hat g_{tt}\,dt^2 \,+\,\frac{d\varrho^2}{\hat g_{tt}}\,+\,\varrho^{2/7}\,d\vec x^{\,2}\,,\\\nonumber\\\nonumber
&& \hat g_{tt}\,=\,\frac{7^2}{\ell^2}\,\,\varrho^2\,\left[1-\left(\frac{\varrho_h}{\varrho}\right)^{\frac{10}{7}}\,\right]\,,
\eea
while the dilaton and the $S^5$ radius remain unchanged. We therefore deduce that the Hawking temperature $T$ and entropy density $s$ for the black brane scale as
\be
T\,=\,\frac{70}{\ell^2}\,\frac{\varrho_h}{4\pi}\,,\qquad\qquad
s\,\sim\, T^{3/7}\, Q^{6/7}\,.
\ee
The dependence of the entropy density on $Q$ is fixed by dimensional analysis. The scaling of the entropy with temperature is also exactly as expected for a theory with Lifshitz scaling symmetry ($s \sim T^{(d-1)/z}$ in $d$ spacetime dimensions).
Taken at face value this implies a vanishing entropy at zero temperature, although for vanishingly small $T$ the classical solution should not be trusted. At any non-zero temperature the dilaton runs down to a non zero value $\propto T^{6/7}$ at the horizon. 

\section{Flow from AdS to Lifshitz}

Finally, we would like to obtain evidence indicating that the Lifshitz-like scaling solution actually emerges as the infrared description of the ${\cal N}=4$ theory with heavy quark density. To show this we would  need to construct a flow connecting the $AdS_5$ (hedgehog) background to the Lifshitz scaling regime. There is a natural dimensionful scale $r\sim Q^{1/3}$, above which the geometry should approach the hedgehog geometry. Below this scale, all terms in the effective potential for the system compete with each other and the system should then enter the Lifshitz-like scaling regime.

\begin{figure}[h]
\begin{center}
\epsfig{file=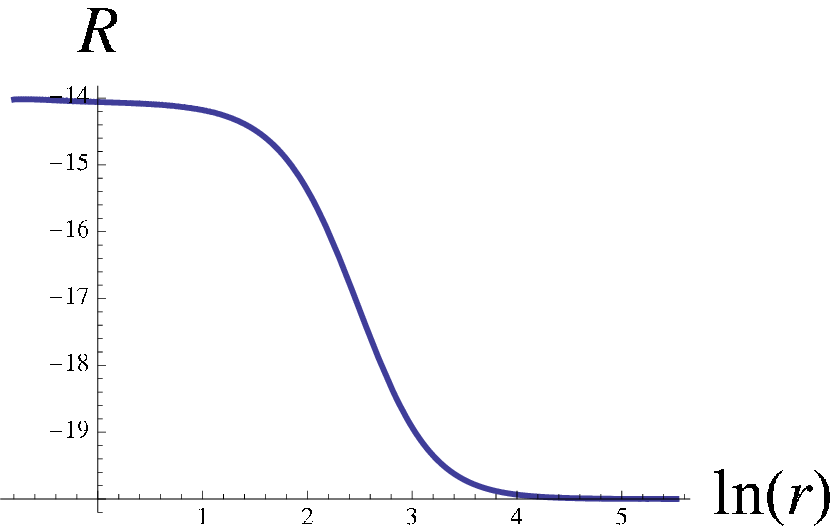, width =2.5in}
\hspace{0.5in}
\epsfig{file=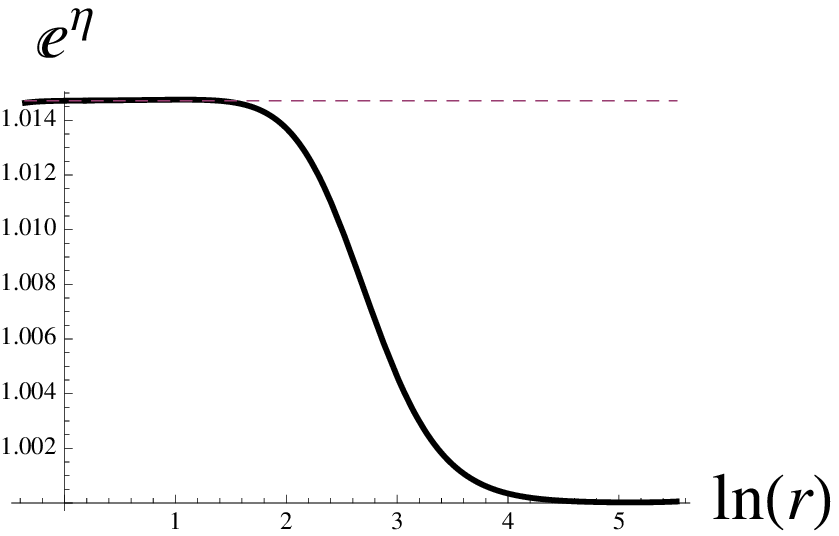, width =2.5in}
\end{center}
\caption{\small{{\bf Left:} The 5D Ricci scalar changes from $152/\ell^2 \simeq -14.06$ in the IR Lifshitz background to $-20$ in the UV, the latter corresponding to $AdS_5$ asymptotics. {\bf Right:} The radius of the $S^5$ flows from unity for large $r$ to $\sqrt[8]{\tfrac{136}{121}}\simeq 1.014$ in the IR.  }}
\label{flow} 
 \end{figure}

\begin{figure}[h]
\begin{center}
\epsfig{file=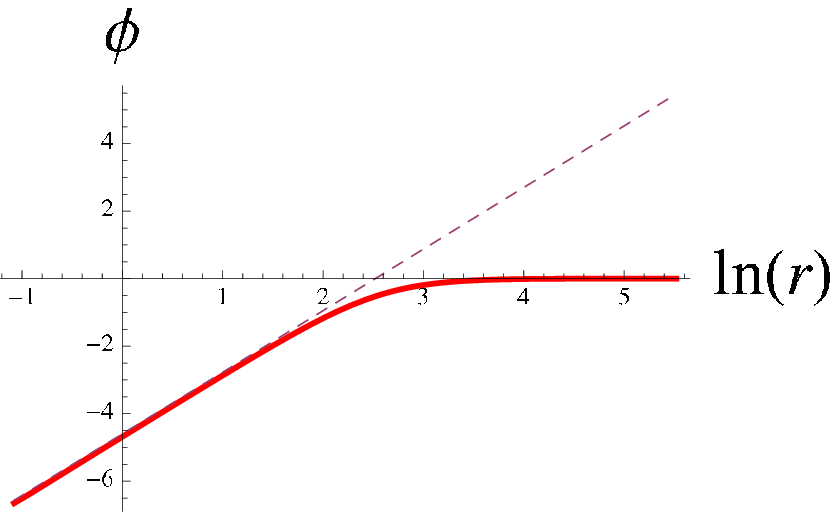, width =2.5in}
\hspace{0.5in}
\epsfig{file=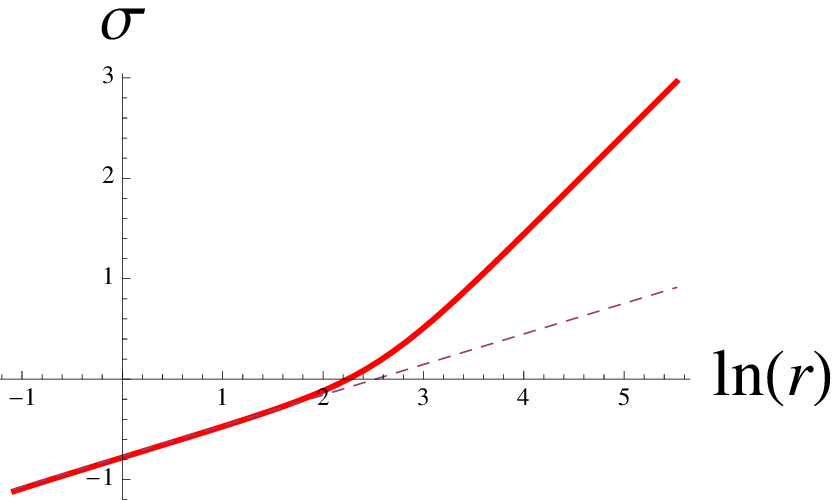, width =2.5in}
\end{center}
\caption{\small{ Working in the gauge $\hat g_{rr}\,=\,{1/r^2}$, the dilaton $e^\phi\sim r^{6/\ell}$, the volume scalar $e^{\hat\sigma}\sim r^{1/\ell}$ and the metric $\hat g_{tt}\sim r^{14/\ell}$ in the scaling regime.
{\bf Left:} The dilaton runs according to the $r^{6/\ell}$ power law in the IR and vanishes in the UV. {\bf Right:} $e^{\hat \sigma}$ obeys the $r^{1/\ell}$ law in the IR and is linear in $r$, as expected in the UV AdS region.}}
\label{flow1} 
 \end{figure}

The natural way of constructing such a flow first requires the identification of relevant and irrelevant deformations of the putative IR background. We have shown that the UV AdS asymptotics \eqref{uv}  does not contain sources for any relevant deformations (preserving the requisite symmetries), although it does allow VEVs for certain local operators. Perturbing the IR solution by an irrelevant deformation and applying a numerical `shooting' method should enable  one to reach the $AdS_5$ asymptotics in the UV. We parametrize the fluctuations $\d_i(\varrho)$ 
about the Lifshitz solution as,
\bea
&&\hat g_{tt}\,=\, 7\ell^{-2}\,\varrho^2\,(1+ \d_1(\varrho))\,,\qquad\qquad \hat \sigma\,=\,\tfrac{1}{7}\,\ln\varrho\,+\, \d_2(\varrho)\,,\\\nonumber\\\nonumber
&&\phi\,=\,\tfrac{6}{7}\,\ln\varrho +\ln\left(\tfrac{11979}{578\sqrt{34} \,Q^2}\right)\,+\,\d_3(\varrho)\,,\qquad\qquad
\eta\,=\,\tfrac{1}{8}\,\ln\left(\tfrac{136}{121}\right)\,+\,\d_4(\varrho)\,.
\eea
At the linearized level, the fluctuations satisfying the equations of motion can each be expressed as a power series, with the leading terms given by,
\bea
\d_1(\varrho)\,\simeq\, a_1\,\varrho^n\,,\qquad\d_2\,\simeq\,a_2\,+a_3\,\varrho^n\,,\qquad
\d_3\,\simeq\,6a_2+ a_4\,\varrho^n\,,\qquad \d_4\,\simeq\, a_5\,\varrho^n\,.
\eea
We find six possible values for $n$,
\be
n_1\,=\, - \frac{10}{7}\,,\qquad n_2\,=\, -1\,,\qquad n_{\pm\pm}\,=\,-\frac{5}{7}\pm\frac{1}{7}\sqrt{\frac{5}{17}\,\left(917\pm8\sqrt{1279}\right)}\,\,,
\ee
where all four sign combinations are allowed. Depending on the values of $n$, the coefficients in the series expansion satisfy special relations and can be identified as relevant and irrelevant perturbations. For each allowed value of $n$, there is precisely one independent coefficient.
The perturbation with $n= -\frac{10}{7}$ is exactly the one that leads to the black brane metric and is relevant in the IR. The $n=-1$ fluctuation corresponds to an additive shift of the radial coordinate. The two perturbations that are {\em irrelevant}  in the IR and can be used to deform the Lifshitz background to alter the UV asymptotics are the ones corresponding to $n= n_{++}=1.973$ and $n=n_{+-}=1.23$. Using these perturbations, and adjusting  their coefficients suitably, we have numerically found a flow interpolating between the Lifshitz scaling regime and $AdS_5 \times S^5$ in the ultraviolet (see Figs. \eqref{flow} and \eqref{flow1}).

The numerical solutions depicted in figs. \eqref{flow} and \eqref{flow1} were obtained in the gauge $\hat g_{rr}=1/r^2$ in which the UV and IR asymptotics take the forms shown in \eqref{uv} and \eqref{sol1}, respectively. The coordinates $r$ and $\varrho$ are related as $r=\varrho^{\ell/7}$. All plots were obtained for $Q=1.89$ and $\phi\to 0$ as $r\to\infty$.
 
In principle, given the numerical flow it should be possible to read off the VEVs, $v_4$ and $v_8$, of the dimension four and dimension eight operators, ${\cal O}_4$ and ${\cal O}_8$ respectively from the UV asymptotics \eqref{uv}. However, we have been unable to accurately isolate these higher order ($\sim r^{-4}$ and $r^{-8}$) terms numerically. There is no reason to expect them to vanish. The fact that there are precisely two independent irrelevant deformations of the IR Lifshitz point is consistent with the presence of the two VEVs in the UV solution. It should be possible to find a relation between the IR deformation parameters and the values of the $v_4$ and $v_8$ 
\footnote{I thank A. Faedo for remarks on this point.}.

\section{Discussion}

The main focus of this work was to understand the influence of a  finite density of heavy quarks on the dynamics of a known gauge theory (namely ${\cal N}=4$ SYM) with a strong coupling string dual. The fact that the system exhibits a flow to a Lifshitz-like scaling regime appears interesting and there are certain aspects of this that deserve further study:

\begin{itemize}
\item{Since the microscopic description of the dual field theory is available, it potentially provides an opportunity to understand, in field theoretic terms, the emergence of the IR scaling regime and the breaking of it by the running dilaton. The insertion of any number of straight timelike Wilson lines, induces a shift in the Lagrangian density,
\be
{\cal \delta L}\,\sim\,\sum_k\delta^3(\vec x-\vec x_{k})\, \bar{Q}_k\left(\partial_t\,-i\,A_0(\vec x, t)-i\,\theta^I_{(k)}\,\phi^I(\vec x,t)\right)Q_k\,\,,
\ee
where $\theta^I_{(k)}$ is a unit vector in ${\mathbb R}^6$ fixing the coupling of the $k$-th quark ``flavour'' to the six scalars in ${\cal N}=4$ SYM (equivalently, the position of a macroscopic string on the $S^5$). The $Q_{(k)}$ can be regarded as one-dimensional fermions living at $\vec x\,=\,\vec x_{k}$ (see e.g. \cite{korchemsky}). Utilizing this as the starting point, it might be possible to shed light on the long-wavelength physics or the ground state of the system, by employing a Hartree type approximation, as in \cite{baryon2}, encoding the collective effect of the heavy quark impurities. }

\item{Perhaps the most natural question following from the gravity picture is whether the IR scaling behaviour is in any sense generic, i.e. can also be shown to arise from similar finite density configurations in other field theories with known string/gravity duals (e.g. ${\cal N}=2$ and ${\cal N}=1$ SCFTs \cite{Klebanov:1998hh} in four dimensions with $AdS_5\times X^5$ duals and the ABJM model in three dimensions \cite{abjm}). A related issue is how the value of the scaling exponent $z$ depends on the microscopic field theory in question, or equivalently, on the corresponding active modes in the gravity dual.}

\item{The smeared configuration considered in this paper preserves  the $SO(6)$ internal global symmetry, but possibly breaks supersymmetry \footnote{We have not checked this, but it appears likely since the individual strings/Wilson lines have different orientations along the internal ${\mathbb R}^6$ directions.}. It should be possible to allow the strings to be localized on $S^5$ (breaking $SO(6)$ to $SO(5)$) while smearing along the spatial directions, and subsequently obtain  backreacted geometries with some unbroken supersymmetry, similarly to \cite{Lunin, DHoker}. }

\item{In the weakly coupled finite $N$ D-brane picture, each individual heavy quark in our setup can be viewed as an F1-string stretching between D3 and D5-branes, with 8 Dirichlet-Neumann boundary conditions. As is well known, such a string behaves as a fermion \cite{baryons, Bachas:1997ui}. It remains to be seen if this fact makes the backreacted smeared string setup relevant for  fermion physics in general, at finite density and strong coupling. }

\item{In the context of the ${\cal N}=4$ theory, the system with smeared heavy quark sources can also be formulated on $S^3$ \cite{inprog}. It is well known that ${\cal N}=4$ SYM on $S^3$ exhibits non-trivial large-$N$ thermodynamics, both at strong and weak coupling \cite{witten, Aharony:2003sx}. The Hawking-Page deconfinement transition at infinite 't Hooft coupling has its  zero coupling analogue which is driven by the Hagedorn growth of states. For theories with fundamental flavours on $S^3$, in the Veneziano large-$N$ limit, continuous deconfinement transitions have been shown to exist at finite quark density, and these are driven by the exponentially large degeneracy ($\sim e^N$) of baryon-like states \cite{Hollowood:2011ep}.  It would be extremely interesting to understand if there is any sign of this physics within the smeared string setup in global $AdS_5$. The hedgehog configurations in global $AdS_5$ in pure gravity, are already known to exhibit a very interesting phase structure in the $T-Q$ plane: a line of first order Hawking-Page transitions ending at a critical point \cite{headrick}. Furthermore, for all values of $Q$, there exist zero temperature black holes with near horizion $AdS_2$ region.}

\item{Another motivation for investigating the above setup on $S^3$ is provided by the IR cutoff given by the radius $L$ of the three-sphere. In the range $L\, Q^{1/3}\gg1$ we may expect the theory to approach the IR Lifshitz-like scaling behaviour, but eventually depart from it due to the finite infrared cutoff. This could be one way to regulate the  deep IR behaviour of the scaling solutions with running dilaton.}

\item{It is not difficult to see that the configuration studied by Headrick in \cite{headrick} also admits a Lifshitz-like scaling solution in the planar case. The goal of that work was to obtain the free energy of ${\cal N}=4$ SYM as a function of the Polyakov loop order parameter i.e. an effective potential for the Polyakov loop at strong coupling. It would be interesting to understand what role, if any, is played by the low temperature scaling regime in obtaining a complete picture of  this effective potential at strong coupling.}

\item{ A key missing ingredient in the heavy quark limit is a $U(1)_B$ global symmetry corresponding to conserved baryon number. As pointed out earlier, this global symmetry and its associated gauge field in the gravity dual will make an appearance once the system is embedded within the framework of ${\cal N}=4$ SYM coupled to a large number of smeared flavour D7-branes giving rise to massive multiplets. For massive flavours, the backreacted geometry in the smeared D7-brane setup of \cite{Bigazzi:2009bk} (see also \cite{Benini:2006hh, Nunez:2010sf}) has an IR geometry (below the quark mass scale) which is simply $AdS_5\times S^5$. Switching on an electric field on the D7-branes (corresponding to a chemical potential/quark density) will result in a spike on the D7-branes behaving precisely like a bundle of smeared strings in $AdS_5\times S^5$. Given the scaling behaviour seen above, it appears likely that embedding the smeared string hedgehog configurations within the D3-D7 setup, will lead to the same IR physics. The presence of a conserved  $U(1)_B$ current would make it possible to define and compute transport properties of the IR theory.}

\end{itemize}

{\bf {Acknowledgements:}} I would like to thank Matt Headrick, Tim Hollowood, Carlos Nu{$\tilde {\rm n}$}ez, Ioannis Papadimitriou, Maurizio Piai and Mukund Rangamani for useful comments and communications. I would also like to specially thank Anton Faedo and Paolo Benincasa for many discussions and for their comments on a draft version of the manuscript. This work is supported by the STFC Rolling Grant ST/G0005006/1.

\end{document}